\documentclass[preprint,11pt,3p]{elsarticle}

\usepackage{amssymb}
\usepackage{amsthm}
\usepackage{lineno,hyperref}
\usepackage{amsmath}
\usepackage{amsthm}
\usepackage{commath}
\usepackage{graphicx}
\usepackage{url}
\usepackage{hyperref}
\usepackage{makecell}
\usepackage{amsfonts}
\usepackage{amssymb}
\usepackage{latexsym}
\usepackage{enumerate}
\usepackage{enumitem}
\usepackage{multirow}
\usepackage{adjustbox}
\usepackage{color}

\journal{ANOR}

\begin{document}

\begin{frontmatter}

\title{The interconnectedness of the economic content in the speeches of the US Presidents}

\author[add1]{Matteo Cinelli}
\ead{matteo.cinelli@infn.roma1.it}

\author[add2,add3]{Valerio Ficcadenti\corref{cor1}}
\ead{ficcadentivalerio@gmail.com}
\author[add2]{Jessica Riccioni}
\ead{j.riccioni@studenti.unimc.it}

\cortext[cor1]{Corresponding author: Valerio Ficcadenti, University of Macerata, Department of Economics and Law. Via Crescimbeni 14 - I-62100, Macerata, Italy. Affiliated to London Centre for Business and Entrepreneurship Research of LSBU (London, UK) from September 2019.}

\address[add1]{Applico Lab, CNR-ISC\\
	Via dei Taurini 19, 00185, Rome, Italy}
\address[add2]{Department of Economics and Law, University of Macerata,\\ Via Giovanni Mario Crescimbeni 
14, 62100, Macerata,Italy}
\address[add3]{London Centre for Business and Entrepreneurship Research, London South Bank University,\\ 103 Borough Road, SE10AA, London, United Kingdom}


\begin{abstract}

The speeches stated by influential politicians can have a decisive impact on the future of a country. In particular, the economic content of such speeches affects the economy of countries and their financial markets. 
For this reason, we examine a novel dataset containing the economic content of 951 speeches stated by 45 US Presidents from George Washington (April 1789) to Donald Trump (February 2017). In doing so, we use an economic glossary carried out by means of text mining techniques. The goal of our study is to examine the structure of significant interconnections within a network obtained from the economic content of presidential speeches. In such a network, nodes are represented by talks and links by values of cosine similarity, the latter computed using the occurrences of the economic terms in the speeches. The resulting network displays a peculiar structure made up of a core (i.e. a set of highly central and densely connected nodes) and a periphery (i.e. a set of non-central and sparsely connected nodes).
The presence of different economic dictionaries employed by the Presidents characterize the core-periphery structure. The Presidents' talks belonging to the network's core share the usage of generic (non-technical) economic locutions like \lq\lq interest\rq\rq{} or \lq\lq trade\rq\rq{}.  While the use of more technical and less frequent terms characterizes the periphery (e.g. \lq\lq yield\rq\rq){}. Furthermore, the speeches close in time share a common economic dictionary. These results together with the economics glossary usages during the US periods of boom and crisis provide unique insights on the economic content relationships among Presidents' speeches.

\end{abstract}


\end{frontmatter}

\section{Introduction}
\label{intro}
Is there a glossary of economic and financial terms whose presence is significant in the US President speeches framework? What is the role of such locutions in the US Presidents public communications? Did the crisis periods affect the choices of mentioning some financial concepts?  Is there a group of US Presidents' speeches that can be classified as a cluster because of the usage of specific terminology devoted do describe economics and financial situations? 

The US President is one of the most influential people in the world. Therefore his communications have to be considered under many perspectives in order to effectively reach strategic objectives. For such a reason Presidents' talks are carefully calibrated on the basis of the audience, the occasions in which the talks occur and the socio-economic surrounding at the moment in which he is speaking. 
The entities of US Society change their expectations, and consequently their actions, also on the bases of the informative set carried out by the Presidents' messages.
Empirical evidence of these phenomena are given by the effects generated by Trump's announcements on tariffs changes for importing the Chinese goods in the US, or the impact of the Obamacare announcements on the US health-care sector. Sometimes, even just tweets can be influential. For example, in \cite{shaban2017event} the authors have analysed about 130 millions of tweets posted during the 2016 US electoral campaign or the studies of tweets occurred during the 2012 electoral campaign presented in {\renewcommand{\&}{and}\cite{maldonado2016twitter}}, and \cite{vargo2014network}.

Since Presidents' talks often contain references to the economic and financial situation of the country 
\citep[e.g.][]{rule2015lexical}, in this paper we aim at checking if words usually devoted to economics and finance make some speeches closer than others along the years. To do so, we have assembled the huge corpora starting from the written version of Presidents talks, from George Washington (April 1789) to Donald Trump (February 2017). In particular, we have used the same process presented in \cite{ficcadenti2019joint} and a summary of it is reported in Section \ref{sec:ecoevaldataset}. The result is a collection of 951 speeches taken from the Miller Center (\url{www.millercenter.org}), a Political Research Institution affiliated to the University of Virginia. We have then investigated the network of the US Presidents' speeches where the nodes are talks and the links are the cosine similarity measures obtained from the frequencies of the economic terms stored at speech level. 

Using text mining techniques that will be widely presented below, we processed the corpora. Specifically, we proceeded to extract the terms contained in speeches whose meaning could be ascribed to the semantic area of economics and finance. These terms result from the merger of two different glossaries: one deriving from a manual used by prestigious newspapers of the sector as The Economics \citep[see][]{bishop2009essential}, and the other is the Wikipedia's glossary of economics see \citep[see][]{wiki:glossary}. After excluding the economics locutions absent in the corpora, we have obtained a total of 383 terms. With them, we have estimated the economic content level of each talk by exploiting their absolute and the relative frequencies {\renewcommand{\&}{and}\citep[see, e.g.][for comparable processes]{wei2015climate, baker2016measuring, tsai2017risk}}.
The usage of network analysis on the speeches dataset through the employment of the frequencies of occurrences of the aforementioned glossary in the talks allows for the measurement of the degree of connections among Presidents, their speeches and their party affiliations.
Hence, the implementation of the network analysis on the speeches of the US Presidents allows to explore the structure of connections and the density of links to understand to which extent different speeches, Presidents and parties are interrelated. The
observation of a clustered structure permits us to differentiate the interactions, the information and the implications deriving from mapping the similarities between the speeches collected.
Our findings are particularly interesting. Especially because we obtain evidence that the historical periods in which the talks have been delivered are relevant for the clustering system, furthermore we detect a core-periphery structure of speeches based on the economic terms commonly used. This means that there is a category of speeches characterized by the utilization of a subclass of economics terms used to describe more specific situations like relevant economics or financial fact. They might be speeches delivered during crisis or post crisis period, or peculiar epochs of reforms.

The structure of the paper is the following: in Section~\ref{sec:lr} we present some comparable works; in Section~\ref{sec:ecoevaldataset} we describe the dataset explaining how it was built and how the economic glossary was created. In Section~\ref{sec:networkmodel} the model for the construction of the network is presented. In Section~\ref{sec:results}, we provide the analysis of the network of US President speeches based on their economic content and in Section \ref{discussion} we discuss the results. Conclusions follow.
  
\section{Literature Review}
\label{sec:lr}
Data mining techniques are used by scientists and researchers to manage massive amounts of heterogeneous data with the aim of extracting useful information \citep[see][ for some recent examples.]{Kocheturov2019, ngai2011application,ravisankar2011detection,malik2018data}
Relevant sub-fields of data mining are text mining and Natural Language Processing. While the former is related to the analysis of a text, the latter  includes natural language comprehension thanks to advanced machine learning techniques \citep[see][]{felici1995talking}. Text mining allows to analyzing strings of characters belonging to texts for extracting relevant information like the meaning of sentences or their sentiment. This is particularly useful when a huge amount of documents is involved or when there is the need of quantifying information from qualitative datasets as in \cite{yuan2018topic}.

Nowadays, textual analysis is employed in a wide variety of studies, for example in medicine \cite{lee2019reviewHepatitis}, in tourism management analysis  {\renewcommand{\&}{and}\cite{cheng2019airbnb}}, in designing recommendation systems \cite{ji2019recommendation}, in analyzing countries' foreign policies \cite{cannon2019shifting}, in investigating the blog users' sentiments during rainstorm and waterlogging disasters \cite{Wu2018} or in understanding the potential applications and users of augmented reality tools \cite{LI2018}. 

Recent contributions like  {\renewcommand{\&}{and}\cite{feuerriegel2018long}} highlight the importance of the information contained in written documents to analyse the economic paths and to forecast economic and financial variables.
Among the latest works, we include {\renewcommand{\&}{and}\cite{feuerriegel2019news}} in which text mining techniques are applied for reducing forecast errors of the macroeconomic indicators by analyzing news.
As part of the prediction of market performance, we mention a study of annual reports of more than a thousand of firms \citep[see][]{balakrishnan2010predictive}, and also the analysis of financial reports through the connection between the words and the financial risk of various companies and banks {\renewcommand{\&}{and}\citep[see][]{tsai2017risk, agarwal2019learning}}.
Furthermore, it is worthy to mention the case of {\renewcommand{\&}{and}\cite{kahveci2016central}}, in which the textual analysis has been employed to explore the semantic  of  monetary  policy  documents  from  the  Federal  Reserve Board,  the  European  Central  Bank  from  2001  to  2015,  and  the  Central  Bank of the Republic of Turkey from 2002 to 2015.

In a more general economic and financial framework we can evidence many works that deepen text mining on the analysis of stock markets, stock returns, trading and,  more generally,  on the ability of transforming qualitative variables into quantitative measures to improve financial forecasts and market predictions {\renewcommand{\&}{and}\citep[see, e.g.][]{blasco2005bad,schumaker2009textual, groth2011intraday, huang2011multilabel, loughran2016textual, mishra2016use}}.
We also mention {\renewcommand{\&}{and}\cite{antweiler2004all, carretta2011impact, hendershott2015institutions, peruzzi2018news}} as examples of studies of the news impact on the financial institutions and on stock market returns; we refer to \cite{nassirtoussi2014text} for a complete review of text mining applied to sock performance predictions.
Within the field of finance, it is also worthy citing some studies in the more specific field of sentiment analysis that has attracted growing interest in the latest years.
Through the analysis of words, punctuation and emoticons collected from social media (such as Twitter), blogs, forums, online reviews, responses to messages from consumers, it is possible to evaluate the positive or negative opinions, their intensity, their emotionalism content and the relevance of the object of analysis with respect to the context {\renewcommand{\&}{and}\citep[see, e.g.][]{tetlock2007giving, loughran2011liability, price2012earnings, garcia2013sentiment, bao2014simultaneously, alfaro2016multi}}. A complete methodological review of text mining applications for financial purposes can be found in {\renewcommand{\&}{and}\cite{kumar2016survey}}, it contains relevant studies with sentiment analysis applications. 
Another interesting review is given by {\renewcommand{\&}{and}\cite{loughran2016textual}}. It contains a summary of text analysis studies in the context of accounting and finance, mentioning pros and cons of each.  Our work does not fall in one of the categories listed in {\renewcommand{\&}{and}\cite{loughran2016textual}}; we present a multi-disciplinary study that shares some common points with dictionary-based information extraction methods and documents/authors classification. Our point of view is different because we want to measure the similarities of the Presidents' speeches, Presidents and Presidents' parties affiliations by using a set of terms recognized as meaningful in the economics and finance field.
The research here presented does not fall in the category of studies where the sentiments are investigated; therefore we avoid all the cons of using pre-trained classification method where the usage of sentiment dictionaries introduces a certain degree of uncertainty concerning the sentiment classes. On the other hand, we use a pre-determined class of locutions and we check their presence into the speeches. In this way, we select the dimensions of the bag-of-word representation considered meaningful for the analysis of speeches closeness under and economics and finance perspective.

Text mining and natural language processing methods are often employed with network analysis. For example, in \cite{chae2015insights}, the authors proposed an analytical framework for analyzing tweets about supply chains and further developing insights into the potential role of Twitter for supply chains practices and researches. Their approach combines three methodologies: descriptive analytics, text mining and sentiment analysis, and network analysis throughout network visualization and metrics. 
The list of papers to be mentioned could be endless, here we have reported few of the most recent studies employing text mining methods and network analysis for economics and finance applications. 

In the context of the US Presidents' communications studies, there are not many works devoted to the analysis of such a wide Presidents' speeches corpus with a combination of text mining and network analysis approaches strictly comparable to our methodological combination. For example, in \cite{light2014words}, the author combined text mining and network analysis to analyse the Presidents' Inaugural Addresses, but he has used the Stanford POS-Tagger and a different similarity measure.
Hence, we use some excellent studies as reference for validating our approach. From a methodological point of view, we are in line with \cite{bail2016combining}. In such a work, the author presents remarkable results in the social field combining text mining and network analysis to study how advocacy organizations stimulate conversation on social media. 
Despite the differences in objectives pursued, our research and \cite{bail2016combining} share the utilization of the frequencies of the words to measure text similarity. Then, both studies involve the usage of communities detection algorithms to capture clusters in the data. In our study, advocacy organizations can be compared to the Presidents, and the speeches can be compared to the posts published by the advocacy organizations. Differently from \cite{bail2016combining}, we computed the cosine similarity to evaluate the distance between the speeches, while Bail has used  the co-presence of terms within the posts on Facebook of advocacy organizations to build a bipartite affiliation network. Moreover, to make our network we have used a prefixed list of locutions whose frequencies of occurrence contribute to the creation of the similarity matrix (as briefly mentioned before, we consider the union of  two glossaries of economics - the one reported in \cite{bishop2009essential} and \citealt{wiki:glossary}'s -  as a tool to explore the speeches proximity).

In our analysis, the main underlying assumption is given by the idea that a set of words related to economics and finance,  whose presence is quite stable during the years (see Figure \ref{economicontent}), can be the core of a common message straighten by the Presidents in their talks. So, we address the meaningfulness of those terms concerning Presidents' speeches connections, to understand the structure of relationships based on such terminology. Furthermore, we look for the presence of a group of speeches based on the presence of such a glossary.  

Another relevant study that inspired our research is \cite{rule2015lexical}, where the authors have performed a wider analysis of the US President speeches but on a sub-sample of our dataset. They have considered the State of the Union Speeches (SoU hereafter) occurred between 1790 and 2014, to investigate changes in topics along the years. Rule at al. have created a set of semantic classes by means of the co-occurrence approaches \citep[][for further info on the method]{callon1991co}, therefore the authors have generated the classes from the speeches (endogenously), and then they have carefully labelled these group on the bases of the main points treated.\\ 
The presence of semantic trends like  \lq\lq Domestic Policy\rq\rq{}, \lq\lq Foreign Policy\rq\rq{} and \lq\lq Political Economy\rq\rq{} is a prerequisite for our analysis. Hence, we start from the idea that the economics and finance are relevant arguments of the political debate, therefore we quantify the relevance of their dictionary in making Presidents' talks closer,  and consequently in creating clusters. Alternatively, from a different perspective, we aim at describing the ability of a class of terms attributable to economics and finance fields to explain the closeness of the speeches, the Presidents and/or their parties affiliations along the years.

Finally, after the determination of the semantic classes, the authors of \cite{rule2015lexical} used a TF-IDF (term frequency–-inverse document frequency) approach to compute the cosine similarity between speeches. They employed a multi-semantic dictionary made by 1000 words whose frequencies are the drivers of the speeches' differences. This procedure is comparable with our even if we decided to use the words' frequencies instead the TF-IDF because we have already selected the terms that matter the most for our study/topic; therefore we do not need an additional weighting scheme.   

Another paper that identifies the words related to economics and finance as remarkable for their ability to divide the media sentiment during the electoral campaign of 2012 is \cite{sudhahar2015automated}. Almost all the words reported in it and related to economics and finance are part of our glossary. The same can be said for the words used in \cite{schonhardt2012yes} to identify economics related part of Reagan's speeches.

In {\renewcommand{\&}{and}\citet{bernauer2009intra}}, the authors used the wordscore approach to measure the intra-party heterogeneity of preferences within parliamentary parties in the German Bundestag during 2002--2005. Wordscore presented in \cite{laver2003extracting} has many common points with our approach, even if it is designed to capture texts' political positions. As for almost all the text mining approach for content analysis, wordscore imposes a priori assumptions about the algorithm training sources. Indeed, it is based on the relative frequencies of words of a pre-selection of corpora that belong to a set of political classes. Therefore, the selection criteria of the set of texts as well as the classes, have to be based on assumptions or on selections suggested by other relevant studies \citep[as stated in][]{laver2003extracting}. Finally, after that a \lq\lq political\rq\rq{} score class-based is assigned to the words appearing into the reference text, the so called \lq\lq virgin\rq\rq{} texts are addressed (in {\renewcommand{\&}{and}\citealt{bernauer2009intra}} the \lq\lq virgin\rq\rq{} texts are those under investigation, whose political orientation is unknown). Basically, the classification of  a virgin text depends on the probability of meeting words with relative frequencies similar to those that appear into one or more reference texts belonging to a certain class. The more the presence of the virgin text words is similar to that appearing in a set of reference texts belonging to a specific class, the more likely the virgin text can be classified as part of that class.

In the operational research field, one of the works analyzing US Presidents political activities is \cite{cochran2014political}. The authors have addressed the US elections using the registered voters' behaviours to determine the best candidate's communication strategy (in term of political positioning) to get their attention.  More in general, operational research studies contain many references to the methods employed to pursue objectives in line with the present study. 
For example, \cite{alfaro2016multi} reports an application of sentiment analysis and opinion mining on comments posted in an organizational and administrative affairs weblog. Similarly to our study, in \cite{alfaro2016multi}, the texts are represented in a document-term matrix and cosine similarity is used to obtain a similarity matrix. But, differently from us, the authors performed the analysis using machine learning techniques as support vector machines and k-Nearest neighbours. In \cite{OLIVA2018}, the authors have developed a novel approach to identify groups of decision makers by means of the coherence of their opinions. They presented the new network based approach by analyzing the situation after the 2012 election. 

Our study differs from the rest of the literature mainly for the peculiar text mining-network based approach employed. The combination of the two techniques allows for an exploration of the data from a topological prospective but keeping into consideration the micro-relationships deriving from the economics words employed. The vector representation of the texts along the economics and finance dimensions (given by the glossary terms frequencies) manifests a great ability in explaining the speeches proximity. Therefore, the analysis of the network highlights clusters and we conclude interpreting information about economics and financial turbulence occurred during US history. Furthermore, the original dataset employed makes this work a unicum for extracting insights about US economy and its risks. 

\section{Data}
\label{sec:ecoevaldataset}

The speeches of US Presidents under analysis in this paper are 951. They have been stated in a period that spans from 1789 to 2017. The transcripts are downloaded from the Miller Center database (\url{https://millercenter.org/}) in June 2017 with a procedure described in \cite{ficcadenti2019joint} and summarized in the next subsection. For each speech we know its date, the name of the speaker and his party affiliation. 

Furthermore, we employ the list of locutions resulting from the union between the glossary of economics resulting in \cite{bishop2009essential}, and the one presented in \cite{wiki:glossary}. The result of this union is appropriately prepared for the analysis object of this paper as we will see in the next section.

\subsection{Data Mining and Pre-processing}
\label{frequencies}

We provide a description of the process realized to make the dataset ready for the analysis.

The first step consists in implementing a web scraping routine to collect the speeches transcripts from the Miller Center website \url{www.millercenter.org}. Furthermore, it is also devoted to code the solutions for managing some common errors occurring at this point due to website inconsistency. For example, we met missed download due to misreported transcripts on the web pages or speeches doubly transcribed into the same page. It implicates memorization of records containing blanks or doubly repeated transcripts.

Secondly, we assessed all the stored transcripts checking for typos. For example, there are missed blanks between words or there are misuses of punctuation creating strings that do not make sense. Therefore, we coded functional solutions to manage them by employing regular expression and removing the punctuation. Once we got rid of these, we used the Hunspell dictionary \citep[see][]{hunspell2017} to check the remaining peculiar words. Indeed, some terms in the speeches are part of ancient English and they are not present in the current Hunspell's dictionary, therefore we have ensured about their existence online and, when they did not result to be typos, we decided to keep them.

We have taken into consideration just the statements that the Presidents have planned to delivery without any external influences provided by journalists or audience interventions. For example, it could happen that the topic of a press conference radically changes after a tricky question from a journalist. This constitutes a deviation from the idea that the President had in mind when he thought about the press conference. For this reason, in the third step, we have eliminated all the transcripts parts that come after the journalists' questions.

The process here summarized in three steps is actually wider, we prefer to refer the reader to \cite{ficcadenti2019joint}, where the process is described with additional details. The refined dataset is made by 951 speeches transcripts containing words stated by all the 45 US Presidents.

As introduced in Section \ref{intro}, we utilized two sets of locutions related to economics and finance to map the speeches in their vector space. In order to do that, we did the following operation:
\begin{equation*}
T = S \cup W
\end{equation*}
where $S$ is the set containing The Economist's glossary of economics whose elements are listed in \cite{bishop2009essential}, $W$ contains the terms listed in \citet{wiki:glossary} and $T$ is the set resulting from the union of the first two. It includes the salient terms employed in a context where economics and finance are treated at a scientific or journalistic level. It comprises names and surnames of relevant economists, but we have decided to replace those bigrams with just the surnames assuming that a President more likely refers to the surname when he wants to speak about a person, especially in official talks. This assumption does not affect the frequencies of occurrence of the bigrams (names and surnames), on the contrary, it allows for a more careful accounting of occurrences. Furthermore, the terms divided by a hyphen (-) are treated as divided, so as bigrams, as well as single words in the hyphened form because there could be different transcription versions (e.g. \lq\lq Most-favoured nation\rq\rq{} has been accounted as \lq\lq Most favoured nation\rq\rq{} as well, see \cite{bishop2009essential} for the definition). Finally, all the locutions originally reported in singular form were pluralized in order to take in consideration both the types; the acronyms have been treated in their extended forms, e.g. \lq\lq OECD\rq\rq{} became \lq\lq Organization for Economic Cooperation and Development\rq\rq{}.

At this point, the number of elements in $T$ is almost doubled with respect to what is resulted from the original union; we grouped the resulting content in different classes on the bases of the number of words that composes the locutions. For example, the word \lq\lq Tax\rq\rq{} has been put into the group of terms made by single words, while the locution \lq\lq Organization for Economic Cooperation and Development\rq\rq{} falls in the set where the expressions are made by six words.  We adopted this subdivision because we want to count the occurrences of the terms in the speeches transcripts without losing or double counting any of them. But, for example, the word \lq\lq Tax\rq\rq{} is contained in locutions like \lq\lq Tax rate\rq\rq{} or \lq\lq Tax avoidance\rq\rq{}. Therefore, if one looks for \lq\lq Tax\rq\rq{} into the speeches transcripts before of looking for \lq\lq Tax rate\rq\rq{}, he/she will introduce a bias because the frequency of occurrence of \lq\lq Tax\rq\rq{} will include the frequency of occurrence of \lq\lq Tax rate\rq\rq{}. To avoid this, we have grouped the locutions on the bases of their length and we got seven groups (indeed, the longest elements in $T$ are made by seven words and the shortest are made by one word). Then, we have looked for the presence of such expressions in the US Presidents speeches firstly accounting for the frequencies of the longest expressions and eliminating them after that. The absolute frequencies of the $T's$ elements are stored in a table and are used to compute the respective relative frequencies at speech level (absolute frequencies dived by speech length for each one). The locutions that do not occur at least once are eliminated and the occurrences of locutions for plural and singular versions have been summed. Finally, the number of elements in $T$ results to be 383. Therefore, the speeches are mapped in a vector space made by 383 dimensions. 

\begin{figure}[!htb]
\centering
\includegraphics{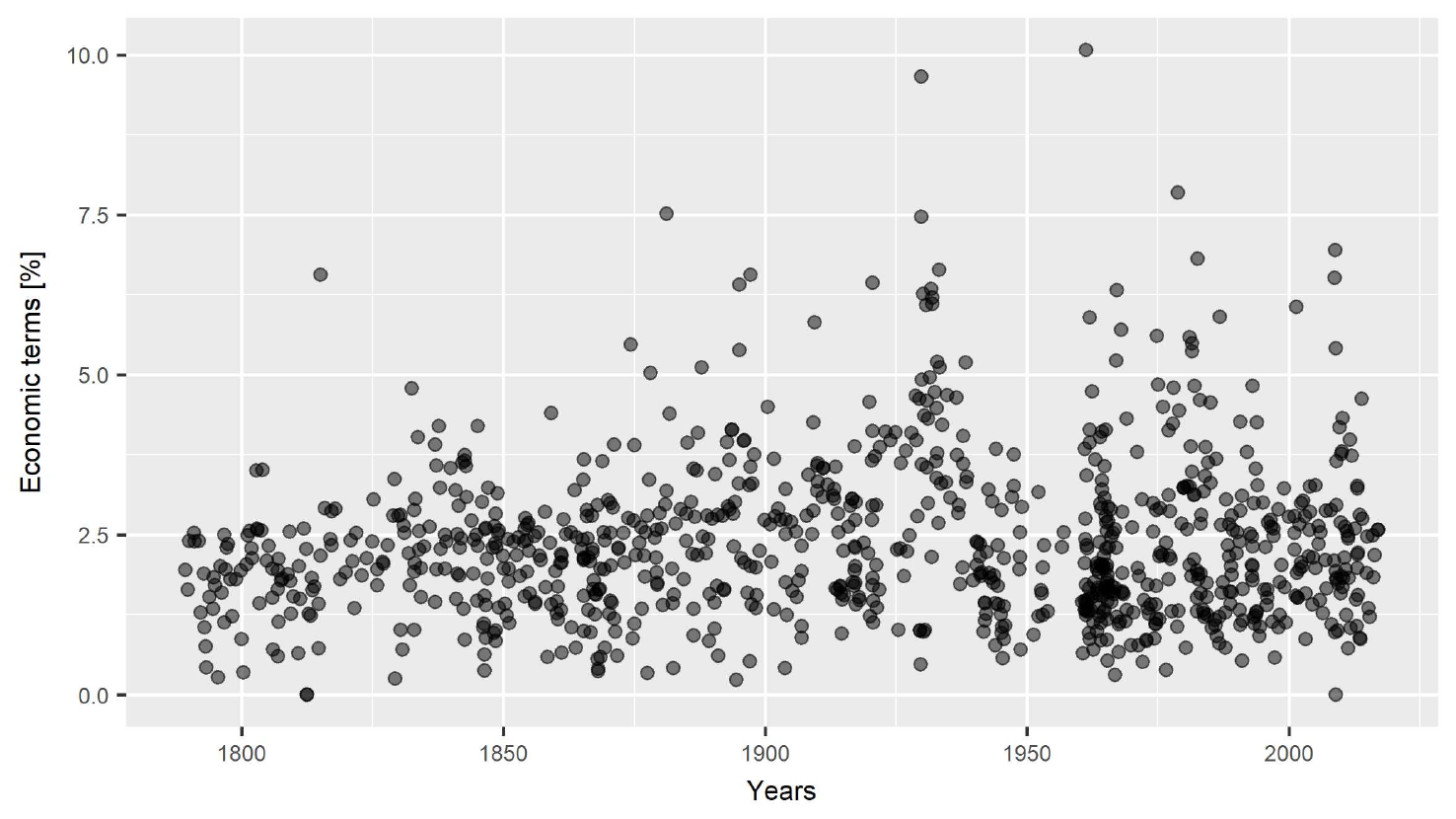}
\caption{Relative frequency of all the economic terms per
speech along the years.} \label{economicontent}
\end{figure}

The sum of the relative frequencies of $T's$ elements for each transcript can be considered as a proxy of the economics and finance content of the speeches. It represents the proportion of a talk devoted to the aforementioned locutions. This measure has a small bias by definition because the frequencies of the economic words might not have one as an upper bound but the effect of this tiny distortion does not affect the key features of the analysis. In Figure \ref{economicontent} each dot exhibits the summed relative frequencies of all the terms belonging to $T$ in the respective speech. It is possible to see that some points fall on zero, they represent three talks where none of the $T's$ locutions occurred:
\begin{itemize}
\item George W.Bush,  Final Press Conference - 12/01/2009 \citep{miller01122009}; 
\item James Madison, Proclamation of Day of Fasting and Prayer -  09/07/1812 \citep{miller09061812}; 
\item James Madison, Proclamation of a State of War with Great Britain - 19/06/1812 \citep{miller19061812}.
\end{itemize}
These speeches have been removed.

Finally, we organize the collected occurrences in a matrix \textbf{I} having 948 rows (one for each speech) and 383 columns (one for each term of the Glossary of Economics terms resulting in $T$). It contains zero when the locution $j$ does not occur into the speech $i$, while it contains the relative frequencies of occurrence if the locution $j$ appears in the speech $i$. 

\section{The network model}
\label{sec:networkmodel}
We describe the system of the US Presidents' speeches as a weighted
complete network $G=(V,E)$, where the $V$ is the set of the nodes
and $E$ collects the edges. The cardinality of $V$ is $n$ while the cardinality of $E$ is $m$.

Each node is a speech. The connection between two speeches $i,j \in
V$ is weighted by the cosine similarity $w_{ij}$:
$$
w_{ij} = \frac {\pmb x_i \cdot \pmb x_j}{||\pmb x_i|| \cdot ||\pmb x_j||}
$$
where $\textbf{\textit{x}}_k$ represents the vector of absolute frequencies of the economic words in the speech $k$, for each $k=1, \dots, n$. 
High values of $w_{ij}$ indicate high similarity in terms of economic content among a couple of speeches, whilst low values of $w_{ij}$ indicate low similarity.
The weights $w_{ij}$ are collected in a matrix $\mathbf{W}$ of dimension 948 by 948. We
assume that $w_{kk}=0$, for each $k$, so that loops are not allowed.

In order to analyse the network, as common practice in networks of correlations such as financial networks~\citep{namaki2011network} and brain networks {\renewcommand{\&}{and}\citep{bullmore2009complex}}, we perform a thresholding of links which are not relevant in terms of strength of association between two speeches.

The thresholding of similarity matrices for filtering out relevant connections employs various methods, somewhat more principled than the use of an arbitrary threshold, whose choice depends on the considered raw data and on eventual information regarding their structure and composition. Such methods are based either on the analysis of the statistical significance of the weights or on the detection of an eventual hierarchical structure of the data. In the first case thresholds derive from analytical arguments (for instance when the underlying data are Gaussian time series), or from permutation tests. While in the second case, methods searching for minimum spanning trees are employed~\citep{battiston2010structure}. For building the network in the case of presidential speeches, we meet a lack of an underlying data structure such as a set of time series and we do not have any (a priori) signals regarding a hierarchical organization of the data. For this reason, to assess the similarity scores contained in $\mathbf{W}$, we statistically test the values $w_{ij}$ by means of permutation tests. We perform a random reshuffle of the elements in each row of the document-term frequency matrix $\mathbf{I}$. Such a reshuffling associates each speech to a randomized set of economic words keeping the distributions of words frequencies and the amount of economic words contained in each speech. Using the reshuffling procedure, a set $I_{rand}$ of 1000 instances of the matrix $\mathbf{I}$ is generated. For each instance in $I_{rand}$, the corresponding matrices $\mathbf{W}$ of cosine values are computed by performing $\binom{n}{2}$ pairwise comparisons among randomized speeches. The resulting set of 1000 cosine matrices is called $W_{rand}$.

\begin{figure}[ht]
    \centering
        \includegraphics[scale = 0.4, trim = 0cm 0cm 0cm 0cm, clip]{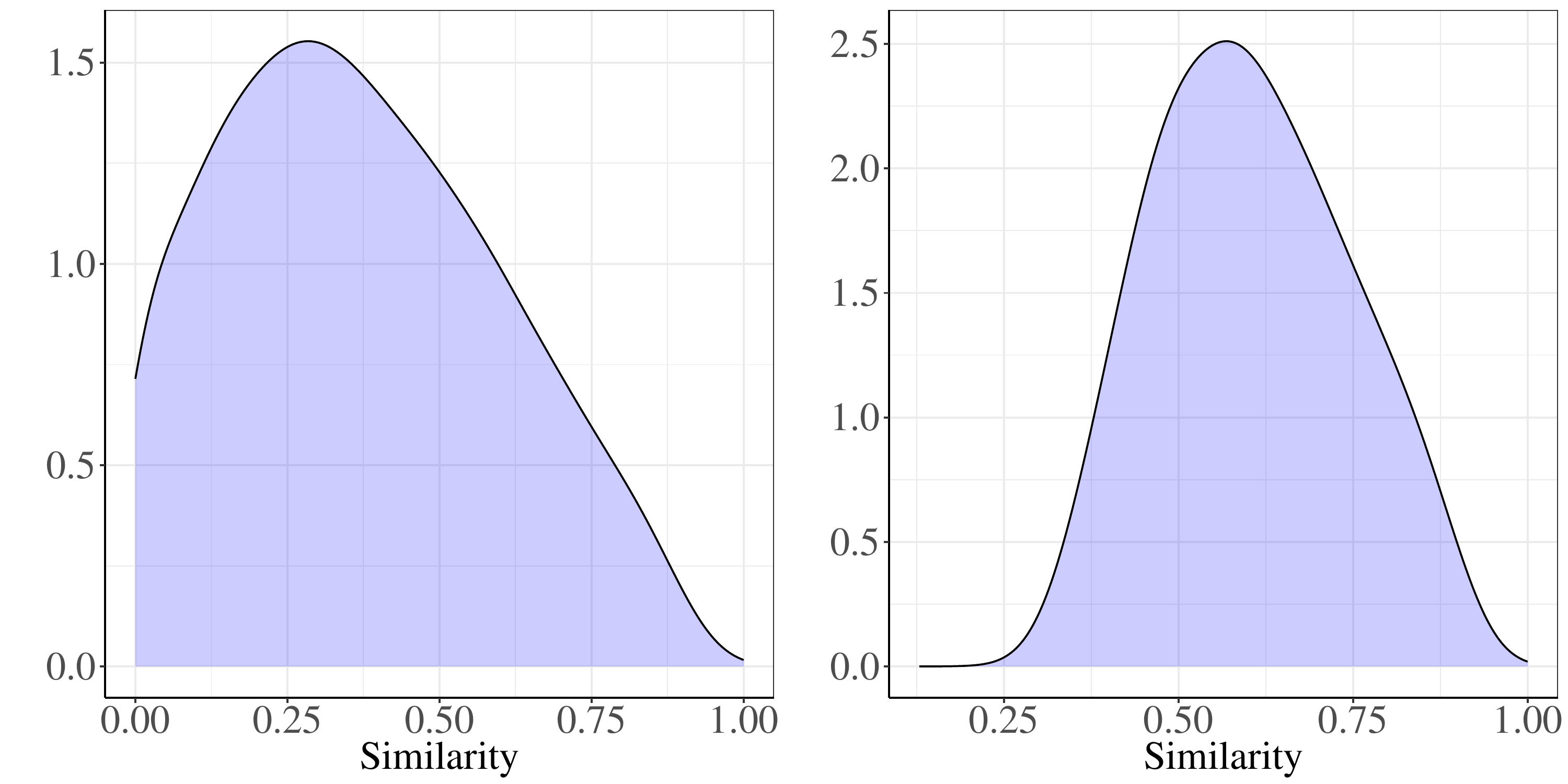}
    \caption{Probability density function (pdf) of cosine similarity of economically relevant words in presidential speeches.
    Left: pdf of the $\binom{n}{2}$ cosine similarity values obtained via pairwise comparisons.
    Right: pdf of statistically significant cosine similarity values. }
    \label{fig:sim_dist}
\end{figure}

Each cosine value in $W_{rand}$ is compared against its counterpart in $\mathbf{W}$ and, for each $ij$, the probability $p_{ij}=p(w_{ij} \in W_{rand} \geq w_{ij} \in \mathbf{W})$ is computed. In other words, for each couple of speeches, we compute the probability that two randomized speeches are more similar to each other than two real speeches. If such a probability results greater than a threshold value $\tau$, i.e. if $p_{ij} \geq \tau$, the corresponding entry, $w_{ij} \in \mathbf{W}$, is discarded since it is not considered statistically significant. The threshold is set to $\tau = \frac{\alpha}{\binom{n}{2}}$ whereas $\alpha=0.001$ and the coefficient $\binom{n}{2}^{-1}$ is the Bonferroni correction for multiple comparisons~\citep{miller1981normal}, that in our case is represented by the number of tested links. The network resulting from such a procedure, in which only statistically significant links are kept, is sometimes called Bonferroni network \citep{tumminello2011statistically}. Additionally, it is worth noting that even if the Bonferroni correction is very conservative the resulting network $G_{\tau}$ is not that different in terms of density, than the network $G_{\alpha}$ that one would obtain without the Bonferroni correction. Indeed, recalling that the network density $d=\frac{2m}{n(n-1)}$, we have $d_{\tau}= 0.342$ and $d_{\alpha}=0.4$.
The empirical distributions of cosine similarity values $G$ (without any corrections) and $G_{\alpha}$ are reported in Figure~\ref{fig:sim_dist}.

The resulting network, after the removal of six nodes disconnected from the largest connected component, has $n=942$ nodes, $m=153677$ links 
and it contains only significant relationships of similarity.
We investigate the structure of such a network in order to understand if the associations among presidential speeches outline a peculiar network structure driven by the presence of the terms belonging to the glossary of economics and finance. 

\section{Results}
\label{sec:results}
\begin{figure}[ht]
    \centering
        \includegraphics[scale = 0.4, trim = 0cm 0cm 0cm 0cm, clip]{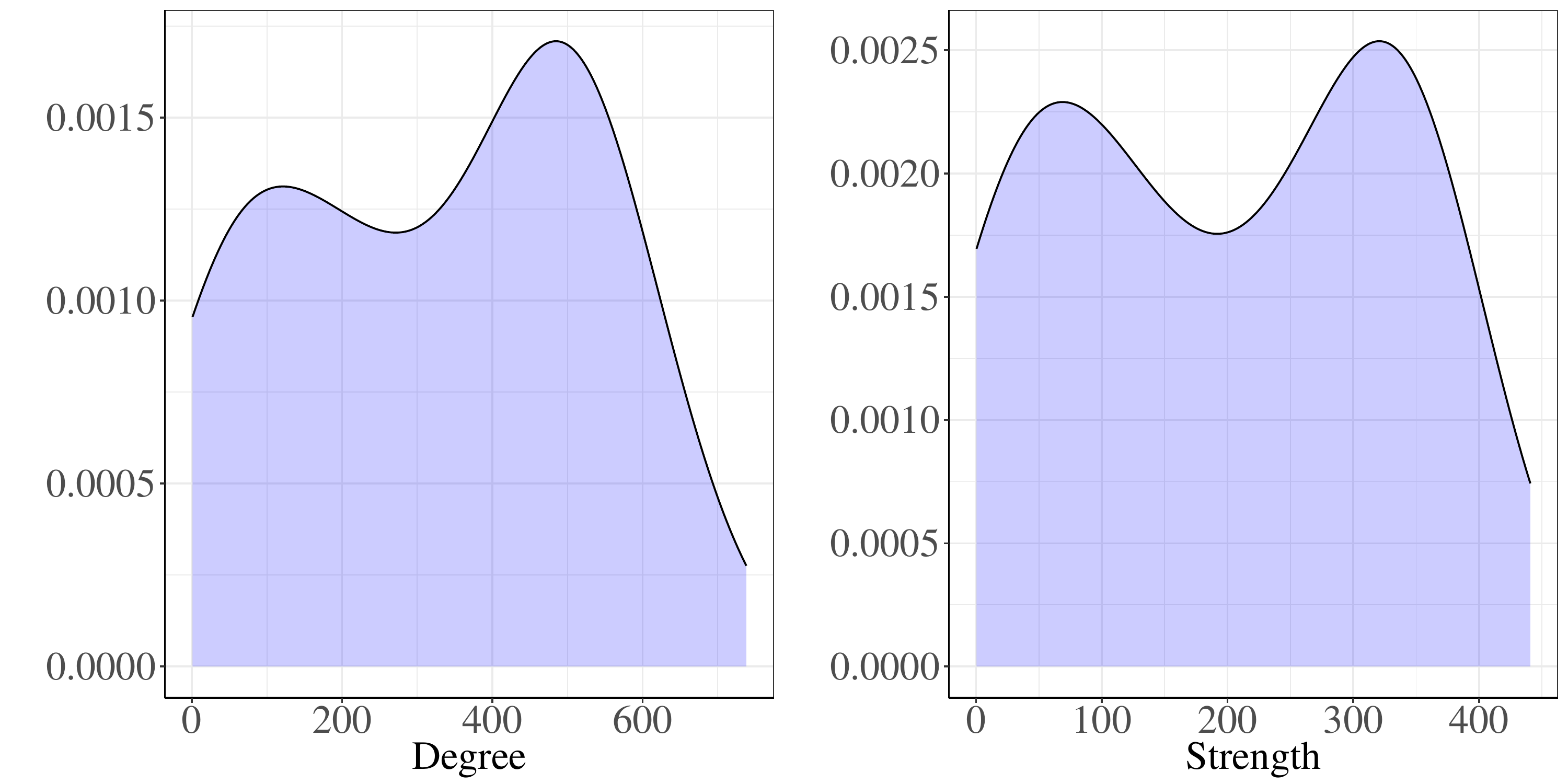}
    \caption{Distribution of degree (left) and strength (right) of the network of presidential speeches. Both distribution display bimodality.}
    \label{fig:ds_dist}
\end{figure}

We firstly analyse the degree and the strength distribution of the network of presidential speeches reported in Figure~\ref{fig:ds_dist}. We observe that both empirical distributions display bimodality, meaning that the nodes could be reasonably partitioned in two different groups characterised by low/high degree and strength.
Another important aspect that we take into account is the clustering of the network, i.e. the cohesiveness of triplets of nodes, by means of both global and local clustering coefficients. The global clustering coefficient $C \in [0,1]$ measures the ratio of closed triangles to connected triples (i.e. subgraphs with three nodes and two or three links). Such a measure of global clustering can be considered both in the case of weighted and unweighted networks {\renewcommand{\&}{and}\citep{opsahl2009clustering}}. The unweighted version of the clustering coefficient can be expressed in the following way:
\begin{equation}
    C = \frac{3\cdot n_\Delta}{n_\land} 
\end{equation}
where $n_\Delta$ and $n_\land$ are respectively the number of triangles and connected triples.
We find that the global clustering coefficient is $C= 0.718$ indicating a relatively high cohesiveness of nodes, surely related to its high density. It also suggests the presence of clusters (communities) in the speeches network.
Such an aspect can be further investigated by considering the local weighted clustering coefficient \citep[see][]{barrat2004architecture} reported in Eq.~\ref{clust_barrat}, that considers the local cohesiveness of each node, combining the topological information with the weights distribution \footnote{It is worth noting the alternatives presented in  \cite{onnela2005intensity} as well as the extended versions of the clustering coefficient in {\renewcommand{\&}{and}\cite{fagiolo2007clustering,clemente2018directed}}.}. 
\begin{equation}
    c_i = \frac{1}{s_i(k_i-1)}\sum_{jh}\frac{(w_{ij}+w_{ih})}{2}a_{ij}a_{ih}a_{jh}
    \label{clust_barrat}
\end{equation}
In Eq.~\ref{clust_barrat}, $w_{ij}$ is the cosine similarity between speeches $i$ and $j$ (as defined in Section \ref{sec:networkmodel}), $s_i=\sum_j w_{ij}$ is the strength of the node $i$ and $k_i = \sum_j a_{ij}$ is the degree of the node $i$. The matrix $\mathbf{A}$, whose elements can be referred as $a_{ij}$, is the binary (unweighted) version of the matrix $\mathbf{W}$.
The local weighted clustering coefficient $c_i \in [0,1]$ groups the structure of the neighbourhood of each node (in terms of connected triplets) with the intensity of connections in the neighbourhood, expressed in terms of links weight.

\begin{figure}[ht]
    \centering
        \includegraphics[scale = 0.4, trim = 0cm 0cm 0cm 0cm, clip]{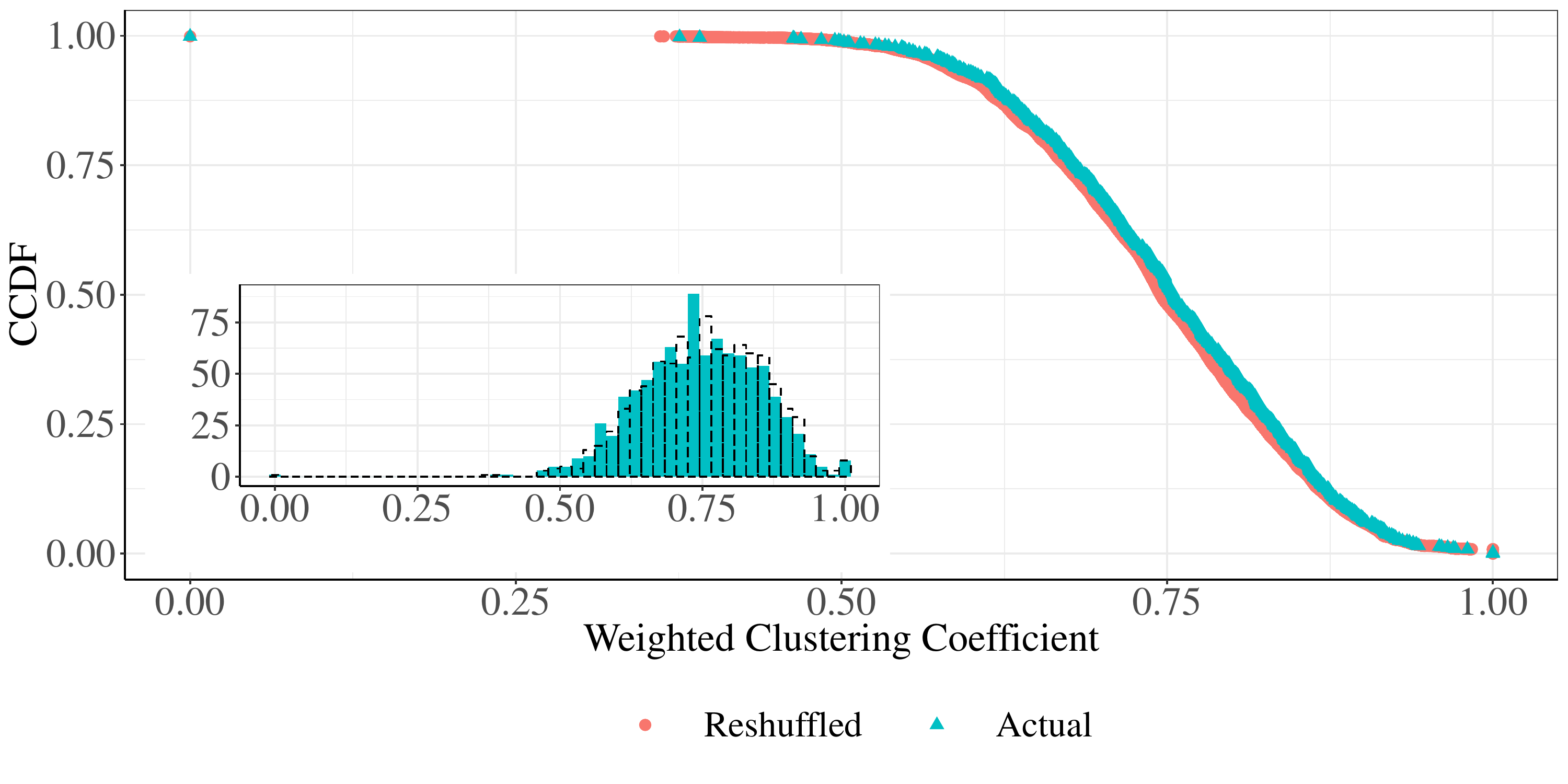}
    \caption{Complementary cumulative distribution function of the weighted local clustering coefficient for the actual network of presidential speeches and for the an ensemble of 100 networks with reshuffled weights and same topology of the actual one. The inset reports the distribution of the local clustering coefficient in its weighted (blue histogram) and unweighted version.}
    \label{loc_cc}
\end{figure}

By plotting the complementary cumulative distribution function (ccdf) of the local weighted clustering coefficient i.e. $P(x>X)=1-P(x \leq X)$, displayed in Figure~\ref{loc_cc}, we observe how a relatively high proportion of the nodes in the network displays a high value of $c_i$ thus indicating the presence of remarkably clustered neighbourhoods. Such an evidence can be associated with a peculiar arrangement of the links weights for two main reasons. First, the average local clustering coefficient in its unweighted form (obtained setting $w_{ij}=cost$ in Eq.~\ref{clust_barrat}) is slightly lower, yet similarly distributed as shown in the inset of Figure~\ref{loc_cc}, than the weighted local clustering coefficient, namely $\overline{c}(w_{ij}=cost)=0.74$ while $\overline{c}=0.75$. Second, the ccdf of the weighted clustering coefficient for the actual network is right-shifted with respect to the ccdf curve of a null distribution associated to such a clustering coefficient obtained from 100 networks with the same topology, but reshuffled edge weights.
Beyond clustering, another interesting quantity is represented by the assortativity coefficient $r \in [-1,1]$ \citep{newman2003mixing} that measures to which extent similar nodes tend to be interconnected relatively to the expected proportion of links under a null model called configuration model. The assortativity coefficient can be computed in the case of scalar node attributes. It can be structural (e.g. the nodes degree) or non-structural (e.g. the year of a certain speech) as in Eq.~\ref{r_scalar},
\begin{equation}
    r=\frac{\sum_{ij}\big(a_{ij}-\frac{k_ik_j}{2m}\big)x_i x_j}{\sum_{ij}\big(a_{ij}x_i^2 - \frac{k_ik_j}{2m} x_i x_j\big)}
    \label{r_scalar}
\end{equation}
where $\textbf{x}$ is the n-sized vector of scalar features whose elements are $x_{i}$ and $x_{j}$, $m$ represents the number of nodes, $k_{i}$ is the degree of the node $i$ and $a_{ij}$ are the elements of the matrix \textbf{A}, binary version of the matrix \textbf{W}. 
The assortativity can also be computed in the case of categorical nodes' attributes (e.g. the political affiliation of the president who gave the speech) as in Eq.~\ref{r_cat}
\begin{equation}
r=\frac{\sum_{ij}\big(a_{ij} - \frac{k_ik_j}{2m}\big)\delta(f_i,f_j)}{2m - \sum_{ij} \frac{k_ik_j}{2m} \delta(f_i,f_j)}
    \label{r_cat}
\end{equation}
where $\textbf{f}$ is the n-sized vector of categorical features and $\delta(f_i,f_j)$ is the Kronecker delta function of the elements $f_i$ and $f_j$ which are components of $\textbf{f}$. Therefore, the assortativity coefficient, similarly to a network-based version of Pearson's correlation coefficient {\renewcommand{\&}{and}\citep{noldus2015assortativity}}, provides us with a value that quantifies the tendency of similar nodes to be interconnected. When we observe a positive value of $r$ we say that the network is assortative, meaning that similar nodes are interconnected while when we observe a negative value of $r$ we say that the network is disassortative meaning that diverse nodes are interconnected.
Interestingly, we could observe very different values of $r$ depending on the attribute that we take into account.

The Presidents' speeches network displays a value of degree assortativity $r_{degree} = 0.04$ indicating no particular mixing to degree while it displays a value of $r_{strength}=0.15$ meaning that nodes with high strength tend to be connected to other high strength nodes while low strength nodes tend to be connected to other low strength ones. Additionally, we observe $r_{date} = 0.177$, $r_{party}=0.009$ and $r_{president}=0.008$ meaning that we observe noticeable associations for what concerns the economics terms presence, among speeches close in time, while we do not observe any particular association among speeches given by Presidents affiliated to the same party or even by the same President.
\begin{figure}[ht]
    \centering
        \includegraphics[scale = 0.4, trim = 0cm 0cm 0cm 0cm, clip]{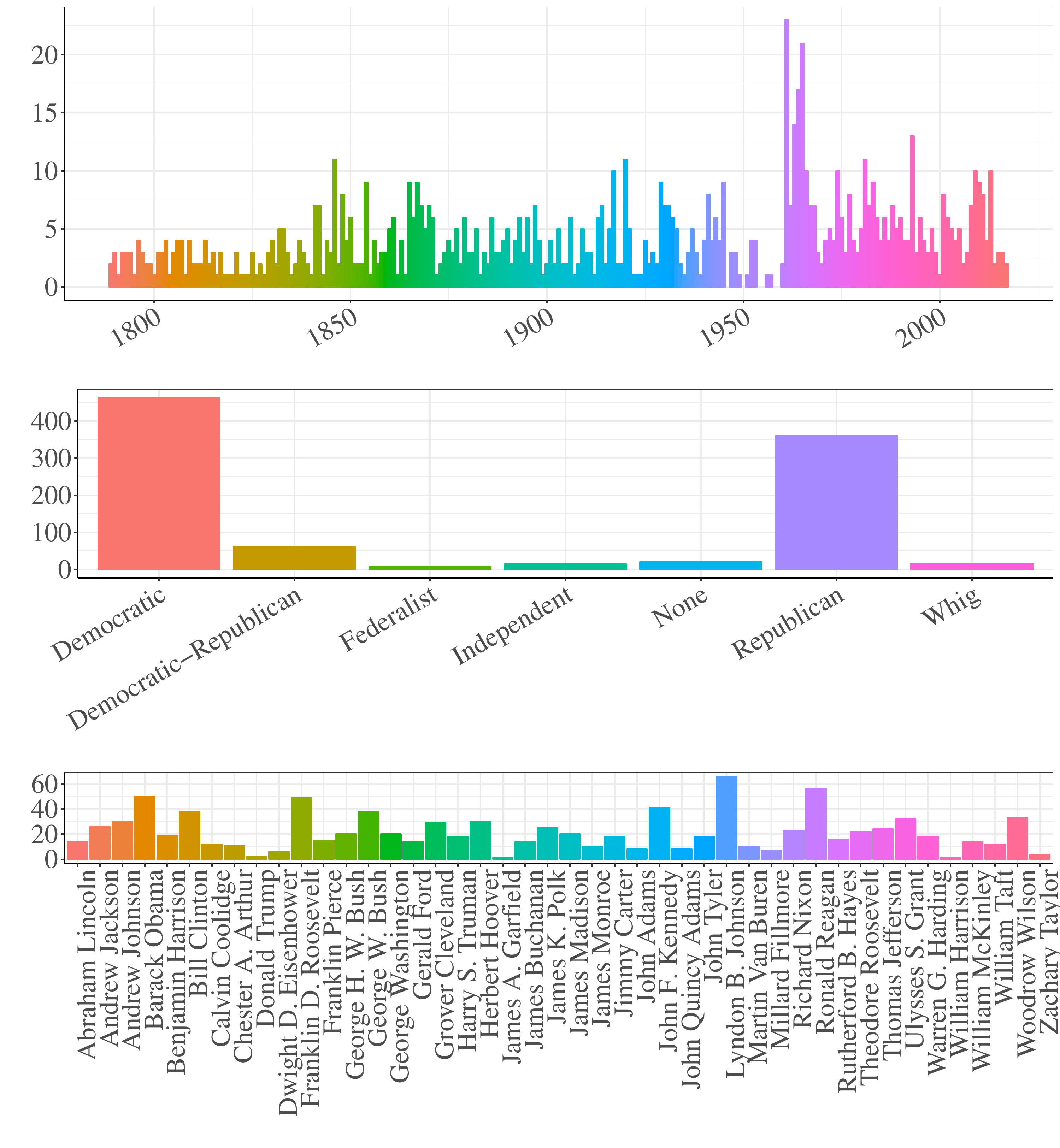}
    \caption{Bar charts of date, party and president's name related to each speech in the network.}
    \label{pn_hist}
\end{figure}
Beyond aspects concerning nodes' attributes, the combination of a high clustering coefficient and a positive strength assortativity coefficient let us room for further investigations related to the presence of communities {\renewcommand{\&}{and}\citep{newman2004finding}} or other higher order framework such as core-periphery structure {\renewcommand{\&}{and}\citep{borgatti2000models}}. 
When the network is divided into communities, we observe subgraphs whose nodes have a higher probability to be linked to the nodes of the subgraph than to any other nodes of the network. When we observe a core-periphery structure, the network topology allows for the partitioning into a set of central and densely connected nodes (the core) and a set of non-central and sparsely connected nodes (the periphery).

\begin{figure}[ht]
    \centering
        \includegraphics[scale = 0.4, trim = 0cm 2.5cm 0cm 2cm, clip]{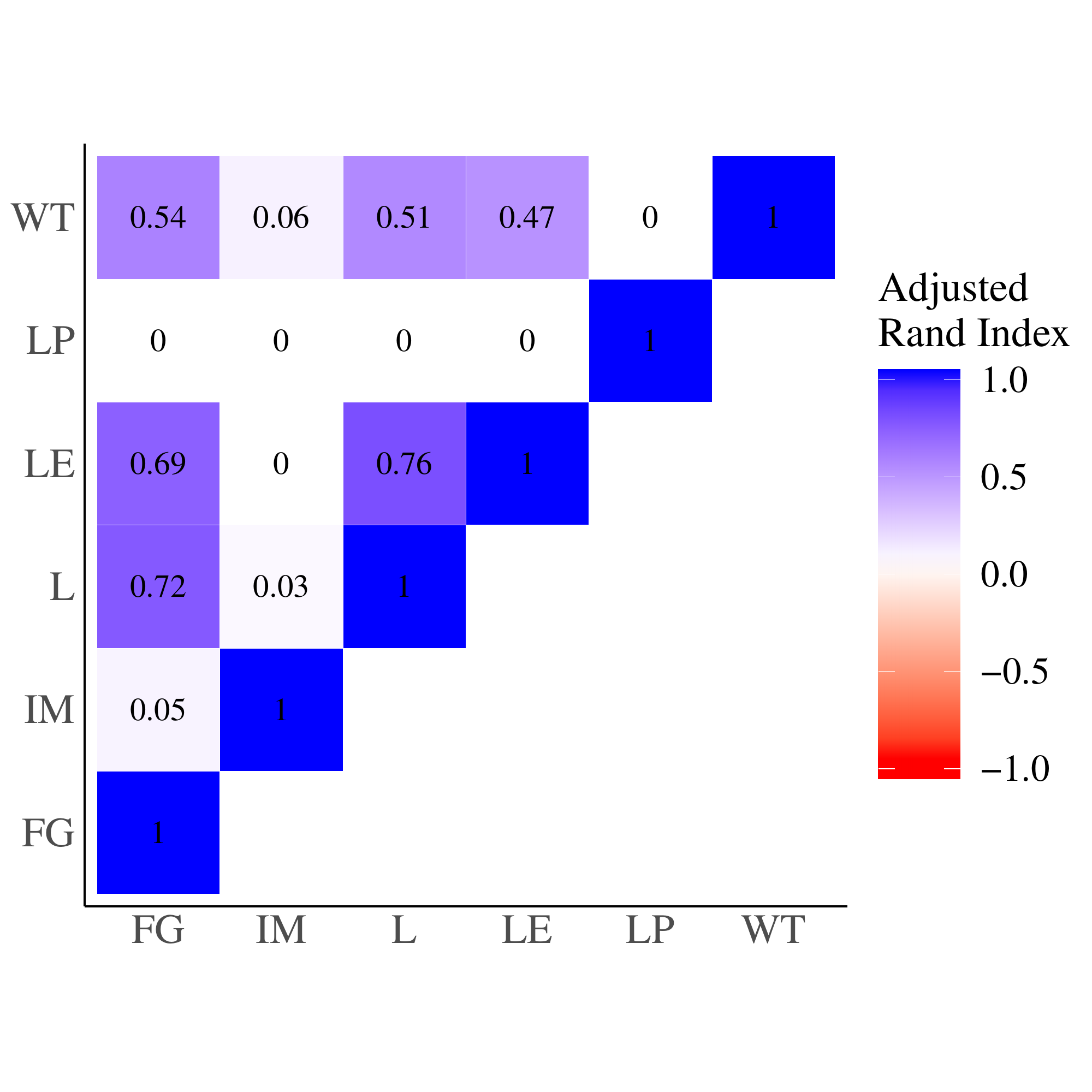}
    \caption{Values of the Adjusted Rand Index for six different methods of community detection.}
    \label{comm}
\end{figure}

We retrieve the community structure of the network by means of community detection algorithms such as Walktrap (WT), {\renewcommand{\&}{and}\cite{pons2005computing}}; label propagation (LP), \cite{raghavan2007near}, spectral partitioning (LE), \cite{newman2006finding}; Louvain (L) algorithm, {\renewcommand{\&}{and}\cite{blondel2008fast}}; Infomap (IM), {\renewcommand{\&}{and}\cite{rosvall2008maps}} and hierarchical aggregation (FG), {\renewcommand{\&}{and}\cite{clauset2004finding}}. We compare the obtained partitions with the Adjusted Rand index $\mathrm{ARI} \in [-1,1]$ {\renewcommand{\&}{and}\citep{hubert1985comparing}}, which measures the agreement of such partitions in terms of assignment of nodes into communities, 
known that $\mathrm{ARI} =1$ indicates perfect agreement and $\mathrm{ARI} = -1$ indicates perfect disagreement. The inconsistency among different values of $\mathrm{ARI}$ displayed in Figure \ref{comm} indicates that there is no overall agreement among the different methods. Therefore, it results in hard to interpret the outcomes. Such a disagreement is also due to the fact that not all the algorithms share the same objective function. For instance IM maximises the so-called map equation {\renewcommand{\&}{and}\citep{rosvall2008maps}} while other algorithms maximise modularity $Q$, a quality function that is analogous to the numerator of Eq.~\ref{r_cat}. Additionally, certain community detection algorithms are normally applied to sparse networks, therefore our outcome may suffer because it comes from the analysis of dense networks.
The evidence from Figure~\ref{comm}, together with the high clustering indicators, assortativity and the slightly bimodal degree and strength distributions suggest the presence of another higher order structure that can further characterize the structure of the considered network. In particular, a positive value of strength assortativity with a noticeable degree of heterogeneity suggests the presence of a core-periphery structure called rich-club {\renewcommand{\&}{and}\citep{zhou2004rich}}. A rich-club is observed in a network when the nodes with the highest degree are tightly interconnected in order to form a dense subgraph. The presence of a rich-club is measured through the rich-club coefficient $\phi(k) \in [0,1]$ which measures the density of the subgraph made of nodes with degree $d > k$. 
The concept of rich-club can be easily extended to measures beyond degree and to weighted networks using appropriate null models for each of the cases, e.g. \cite{opsahl2008prominence, cinelli2018rich, cinelli2019generalized}.
In this one, we assess rich-club ordering in the case of node strength and we measure the density of connections among nodes with the highest strength $\phi(s)$. The value of $\phi(s)$ is compared against its average value $\Bar{\phi}(s)$ across an ensemble of 100 networks with the same topology but reshuffled edges weights. When the ratio $\phi(s)_{norm}=\frac{\phi(s)}{\Bar{\phi}(s)}>1$ the network is said to display rich-club ordering.
It is worth noting that the null model that we are taking into account preserves the topology while reshuffling the edges weights; it means that the nodes strength distribution is not preserved across randomized networks. For this reason, as explained in~\cite{cinelli2019generalized}, we rank the nodes by increasing strength in each of the considered networks and we measure the density of connections among nodes whose rank is higher than a value $p \in [1,n]$. Accordingly to the explanation given for $\phi(s)$ we computed the index $\phi(p)$ and its normalized version $\phi_{norm}(p)$.

By computing the rich-club coefficient for each value of $p$, we obtain a curve of the density of the considered subgraph which equals one when such graph is complete. 

%
\begin{figure}[ht]
    \centering
        \includegraphics[scale = 0.4, trim = 0cm 0cm 0cm 0cm, clip]{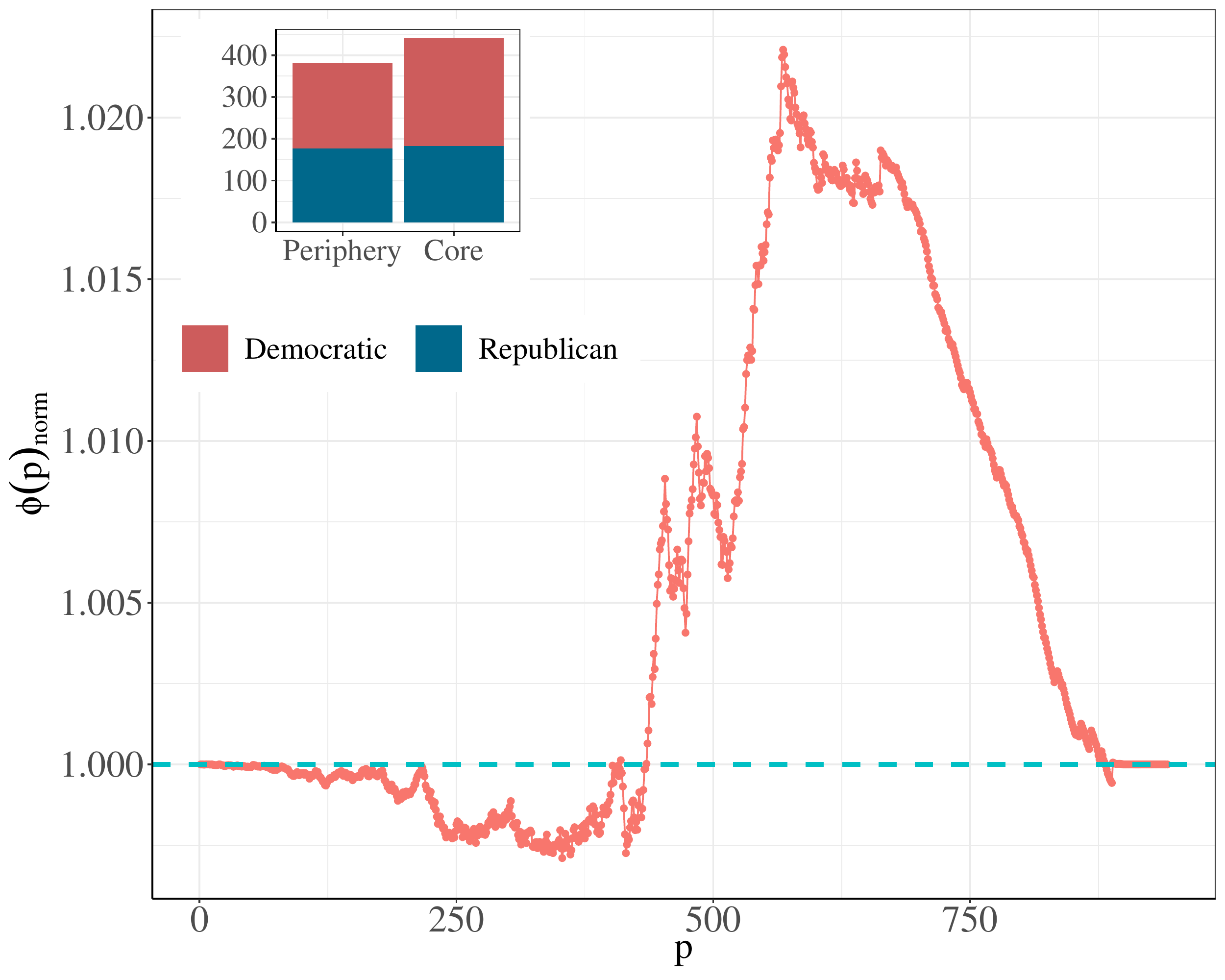}
    \caption{Normalized rich-club coefficient of the presidential speeches network. Rich-club ordering is observed for values of $\phi(p)_{norm}>1$. The inset reports the number of speeches inside and outside the rich-club considering also considering the affiliation of the speaker. For the sake of representation only affiliations to Democratic and Republicans have been reported. }
    \label{rc}
\end{figure}

From Figure~\ref{rc} we observe that the network of presidential speeches displays a rich-club ordering and that it enters in the so-called rich-club regime (i.e. $\phi(p)_{norm}>1$) for a value of $p \sim 430$. Additionally, by plotting the amount of speeches given by  Presidents affiliated to the two main parties in the US, we note how the network nodes, both inside and outside the rich-club, display a balanced proportion of political affiliations, meaning that the partition of speeches tends to resemble the economic content more than other political aspects. 

\begin{figure}[ht] 
  \begin{minipage}[b]{0.5\linewidth}
    \centering
    \includegraphics[width=1.2\linewidth, trim = 2cm 3cm 0cm 3cm, clip]{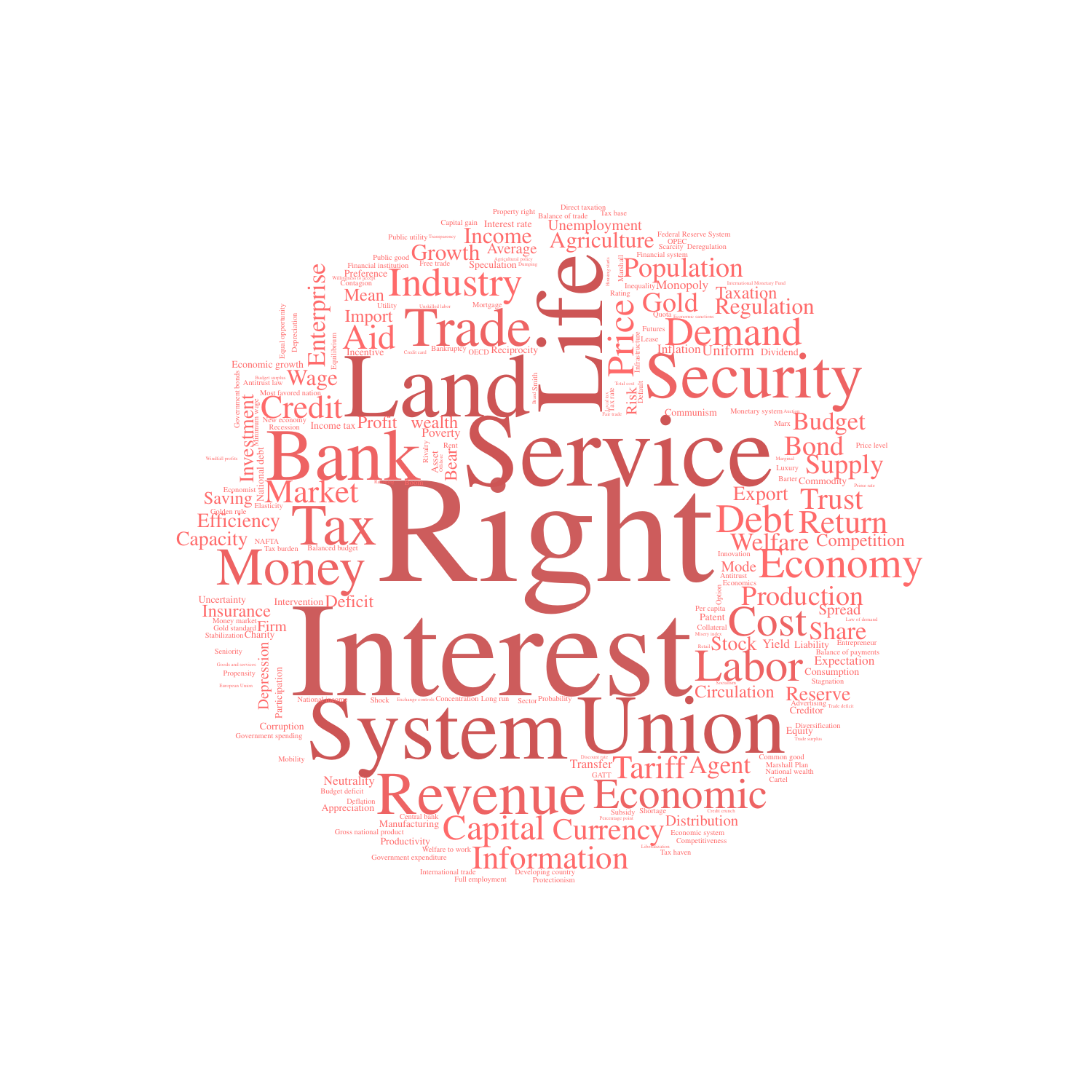} 
    \vspace{4ex}
  \end{minipage}
  \begin{minipage}[b]{0.5\linewidth}
    \centering
    \includegraphics[width=1.2\linewidth, trim = 2cm 3cm 0cm 3cm, clip]{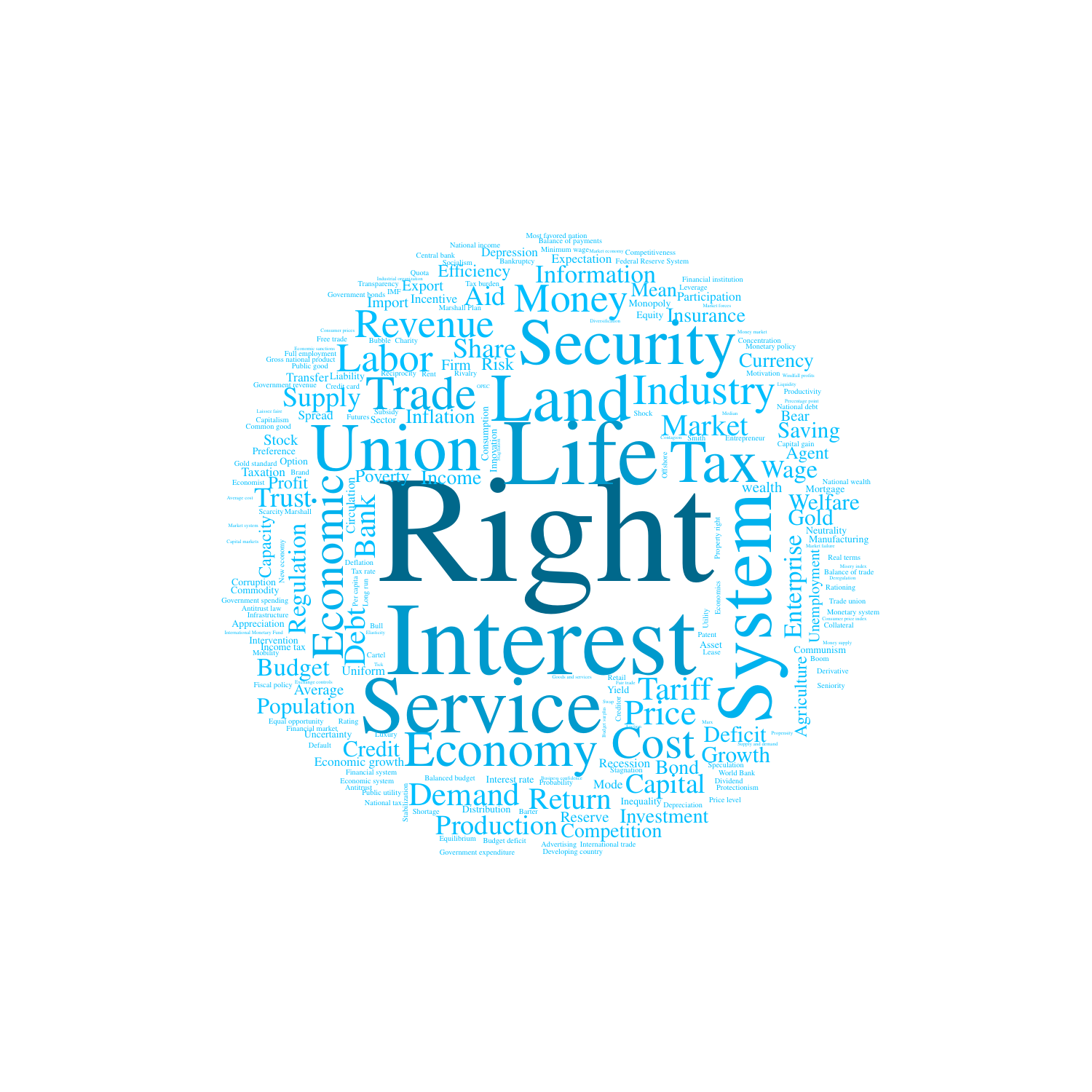} 
    \vspace{4ex}
  \end{minipage} 
  \begin{minipage}[b]{0.5\linewidth}
    \centering
    \includegraphics[width=0.8\linewidth]{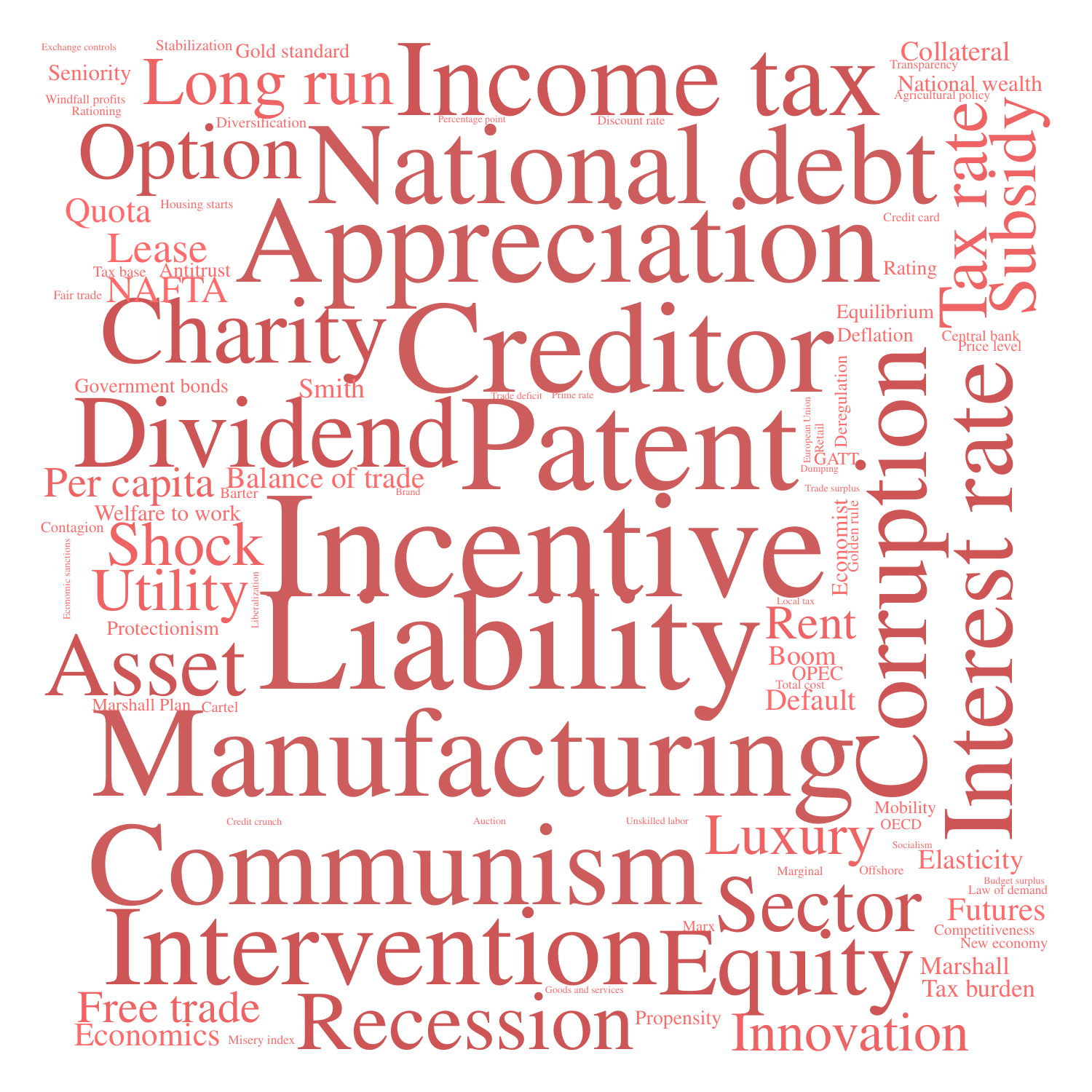} 
    \vspace{4ex}
  \end{minipage}
  \begin{minipage}[b]{0.5\linewidth}
    \centering
    \includegraphics[width=0.8\linewidth]{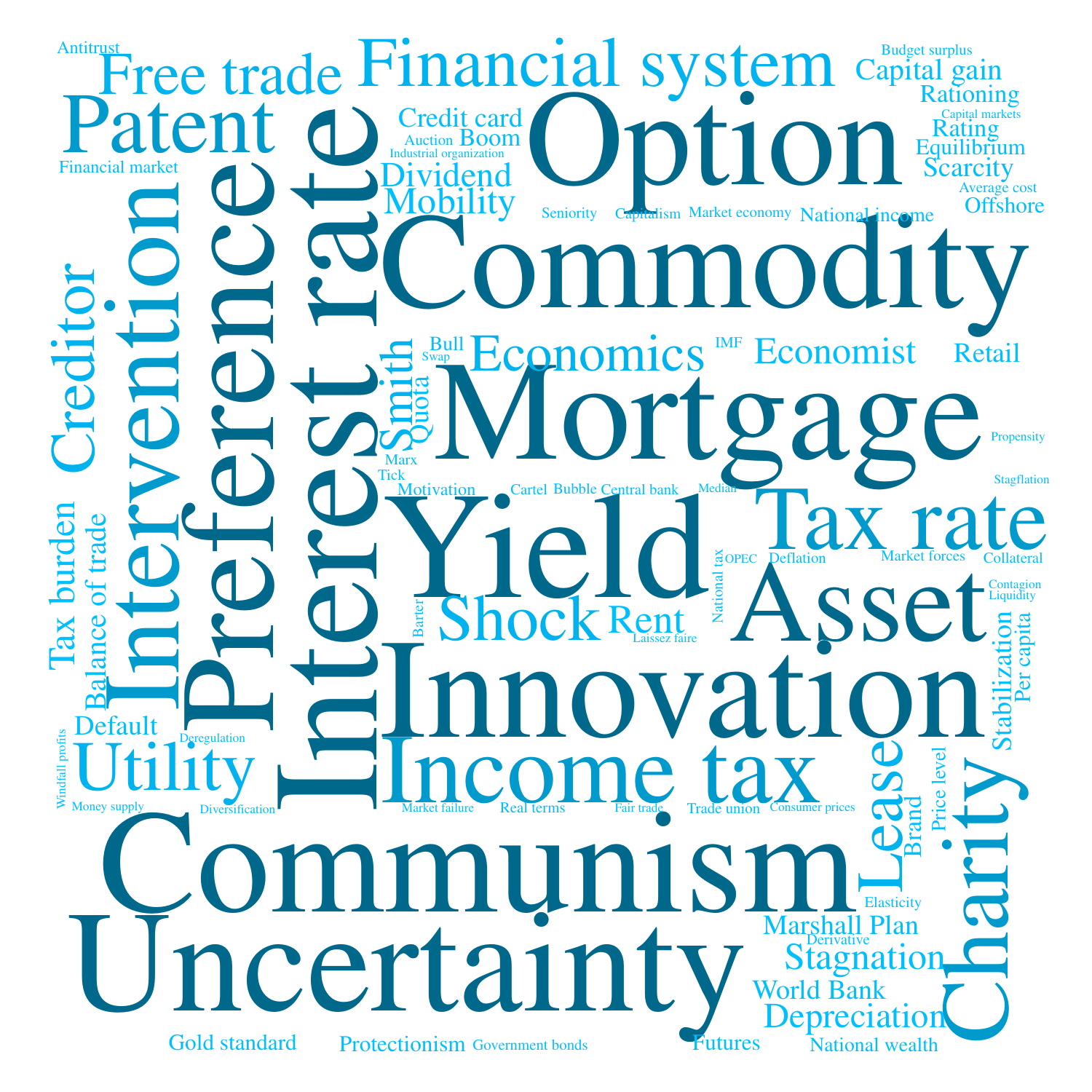} 
    \vspace{4ex}
  \end{minipage}
  \caption{Wordcloud for the rich-club (left) and the rest of the network (right) obtained considering two partitions of the matrix I. The size of the words is proportional to their frequency. In the top panels we consider the entire set of words while in the bottom panels we cut out words with very high frequencies. We observe that while the main economic concepts are common to both the rich-club and the rest of the network, less frequent concepts differ between the two groups. }
    \label{fig:wordclouds}
\end{figure}

\section{Discussion}
\label{discussion}

Our analysis shows the relevance of the terms belonging to the glossary of economics and finance in connecting the Presidents' speeches between each other. We employ a text mining-network analysis approach to measure the similarity of the speeches by using the aforementioned glossary.

In Section \ref{sec:results} we note high levels of clustering coefficients at local and global level that indicate the capacity of the speeches of being organized in communities. The relationships between speeches are quite complete thanks to the presence of many triangular schemes. This aspect confirms the idea that the analysed glossary of economics is recurrent in the political debate and it constitutes a significant component of the connection between Presidents' talks. Looking at Figure~\ref{economicontent}, one can find further confirmations of this fact by noticing the high density of the points in certain periods as well as the quite stable presence of the economics locutions along the years. The talks interconnections and the presence of local triangles bridged with other closed triangular relationships are indications of the use of a shared terminology within local communities of speeches. It might be related to the type of phenomena faced by the speakers during different socio-economic periods (e.g. recession periods characterized by high unemployment rate or crises leaded by particular sector of the economy).

\begin{figure}[ht]
    \centering
        \includegraphics[width=\textwidth]{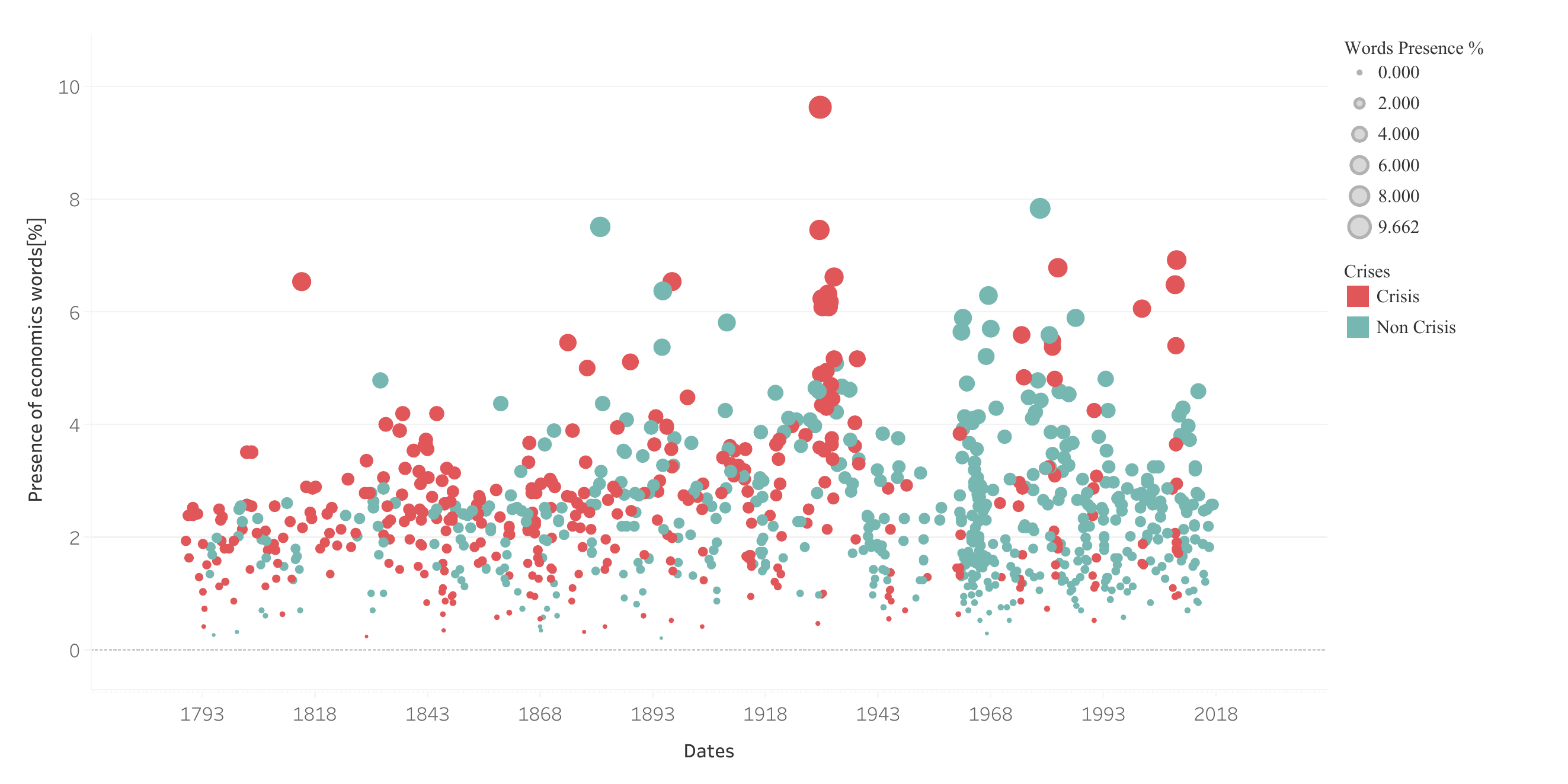}
    \caption{Percentage of the economics terms per
speech along the years divided by period of recession. Such a division has been taken from the Wikipedia's list of crises, see \cite{wikicrisis}. The red indicates that the speech has been stated during a period of recession, while blue indicates a non recession period}
    \label{crisisdummy}
\end{figure}

The assortativity analysis allows for the determination of the features that can be relevant in justifying the relationships between talks. The structural assortativity, measured by using the strength of the nodes, reveals that speeches characterized by close levels of similarity tend to be closer. This offers further hints on the existence of sets of talks that share the utilization of some economics terms picked from our glossary. They might be the State of the Union or the Inaugural Addresses for example. During these type of talks, the Presidents usually devote part of their speech in commenting the economic situation of the country, sometimes involving a more specific/technical analysis of the indicators \citep[see][]{rule2015lexical, light2014words}.
We check for the influence of the political parties affiliations of the Presidents, the years in which the speeches have been stated and the speakers as discriminant features to identify the clusters. The strongest contribution is provided by the years of the talks, indeed the assortativity for that case manifests noticeable relationships between talks belonging to the same years. Namely, public communications that are close in time tend to be similar for what concerns the employed economics terms. As an example, the devastating crisis of 1929 was the starting point of a turbulent period of high instability and the US economy and financial conditions have been at the center of the debate for long time. Indeed, during the years 1929 - 1933 the Presidents have spoken a lot about the crises (see Figure \ref{crisisdummy}), therefore we expect speeches occurred at that time to be closer with respect to those stated during war periods. 
Another example is given by the more recent financial crisis started in 2008 (see Figure \ref{crisisdummy}).\\ 
The low level of assortativity that emerges when we test the contribution of Presidents (so the relevance of the speakers in clustering the speeches) and their affiliations to parties are partially justified by a low level of politicization in the usage of economics and financial terminology. Furthermore, the Presidents usually utilize the political rhetoric to accentuate the government difficulties coming from exogenous factor to exalt their achievements or to manifest awareness of the country challenges. This is particularly true for Presidents that have faced crises as well as for the Presidents that had the responsibility of leading the US after a tough period. This phenomenon generates reference to the economy and finance in the US Presidents' speeches.\\
In addition, if one looks at changes of governments occurred during or immediately after the toughest US crises, e.g. \textit{Great Depression} and the \textit{Great Recession}, it is possible find additional justifications for the low levels of assortativity by Presidents and their affiliations. Indeed, these two crises have involved the efforts of more than one government and President. Specifically, the \textit{Great Depression} occurred during the Republican Presidency of Hoover has produced aftermath during the successive Presidency of the Democrat Roosevelt. While the \textit{Great Recession} has first involved the Republican Bush and then the Democrat Obama. It means that, during the period in which there are picks of economics words usages (crisis or booms periods for example), economics and finance are addressed by different Presidents and therefore different parties. 
Furthermore, it confirms that during a mandate, the way of using economics terms is common among Presidents and parties.\\
Certain events have such a disruptive power that the whole society pays attention to them. Therefore, in speaking about these facts, the parties division does not matter; the public debate embeds such events regardless of the storytellers' identity. Moreover, the President is the most important political landmark in the US, therefore each time something relevant happens in the country, he has a sort of institutional obligation to speak about it, regardless his political affiliation or his sensibility. Consequently, the lexicon used by different Presidents does not change much, especially considering that our glossary is made up of a terminology useful to identify concepts (nouns for example) more than sentiments or adjectives associated to them.\\
 The talks particularly devoted to economics and finance are probably characterized by the contribution of technicians and ghostwriters with high knowledge of the economy. It means that the Presidents attitude are not so manifested in these cases. In addition, as said, the glossary here used does not allow to capture contributions coming from ideologies or different visions of the economic system. This is expected to be an important feature for discriminating the speeches on the bases of the parties for example.\\
Ultimately, the assortativity by party can be affected by the unbalanced proportion of speaker affiliations to different political parties (we have more speeches stated by democrats than republicans, see Figure~\ref{pn_hist} - mid). While the distributions of speeches by President and by dates are relatively homogeneous, as displayed in Figure~\ref{pn_hist} (bottom and top).
However, as suggested in~\cite{newman2003mixing}, we have to carefully consider these results given that in networks with many attributes, disassortative mixing tends to resemble a random assignment of node attributes since, in presence of several attributes, then random mixing will most often pair unlike nodes.

The analysed network results to be relatively dense despite the threshold applied to links for avoiding scarce statistical significant links. Such density value reflect the idea of having a glossary of economics particularly popular among the Presidents. But, at the same time, it may create noise when community detection algorithms are applied (see the disagreement between methodologies in Figure \ref{comm}).

The research of a core-periphery structure grounds on the positive and remarkable assortativity based on nodes' strength and on the bimodal distribution of the nodes' degree and strength. Therefore, we look for a core of strongly interconnected speeches in which there is a regular presence of common terms referred to economics and finance. 
Figure~\ref{rc} confirms the presence of a rich-club (when $\phi(p)_{norm}>1$) for $p \sim 430$. Consequently, we can state that the discriminant between the two groups should be reflected by the presence of two sets of economics terms, which drive the differences of similarity regime. They can be represented by words clouds as displayed in Figure~\ref{fig:wordclouds}. The two clouds at the top of the figure show similar words occurrences (dimension of the font), therefore the nodes' strength within the core and the periphery has to be driven by the presence of locutions reported in the two clouds shown at the bottom of Figure~\ref{fig:wordclouds}. Indeed, the bottom clouds contain the words less frequently occurred within the two groups. Namely, the core and the periphery share a common set of words and the occurrence of more marginal terms conditions the belonging to the core or to the periphery of a speech.  

Concluding, the network results to be divided into a core of speeches where nodes with the highest strength are connected to each other and a periphery with the opposite characteristic. This outcome implicates the aforementioned discriminatory behaviour of some terms' frequency distribution. So, given a core of words regularly present among the speeches, the dimensions of the vector space devoted to bring/eject a speech in/from the rich-club community is provided by a relatively high presence of terms characterizing a contingent situation in which the talk has been stated. For example, the word \lq\lq Manufacturing\rq\rq{} has a notable presence in the set made by speeches belonging to the rich-club but, in contrast, it disappears from the rest of the network (bottom Right cloud of Figure \ref{fig:wordclouds}). On the other hand, the bottom blue cloud shows the presence of \lq\lq Innovation\rq\rq{} in the peripheral network but not in the core (bottom left cloud of Figure \ref{fig:wordclouds}).
 
 The core-periphery structure can be interpreted under the prospective shown in Figure \ref{crisisdummy}. Namely, by comparing the percentage of economics words present into the speeches stated during recession periods versus the presence of such words in the talk delivered during the rest of the time \citep[see][for the classification of crisis period in the US history]{wikicrisis}. Figure \ref{crisisdummy} clearly shows two regimes with a  change between 1940 and 1960. The number of crisis has diminished during the last century, therefore the number of speeches delivered during the recession has slowed down. Furthermore, by looking at the last crisis (2008 - \textit{Great Depression}), it is possible to notice few big red points. The fact that during the last crisis the Presidents have used fewer economics terminology than before can be attributed to the need of modern leaders of avoiding media identification with turbulent periods to maintain low association with unpleasant events. However, the red points manifest lower presence of economics words and seems to be located in the past, while the blue points are more concentrated during recent years and they seem to have a higher mean. This has certainty conditioned the similarity, contributing to the creation of a core-periphery structure.
 
\section{Conclusions}
\label{conclusions}
In this study we have designed a procedure made by a combination of text mining and network analysis techniques. The tools employed to perform the analysis are well established in the literature (see Section \ref{sec:lr}) but their combinations and the application in this context constitutes - at best of our knowledge - a unique case. The designed approach is general and we believe that the usage of it in other fields can provide promising results.

Here we focused on economics and finance topics that are recurring in political discussions. Indeed, Politicians have a preponderant role in deciding the fiscal and monetary policy; their actions impact the economy of the country as well as the financial sector. The Presidents' public communications related to economics and finance are designed to reach certain listeners and to spread different type of messages on the bases of contingent situations, the audience and political objectives to be reached. Anyway, the 45 Presidents have referred to the economy and financial sector at least once in their political life. \\
This study highlights the relevance of the economic jargon employed by the US Presidents since the foundation of the country. Indeed we have found a quite stable presence of the terms belonging to the glossary of economics in the talks; see Figure \ref{economicontent} for a visual inspection of this fact. \\
The network analysis approach allows to detect communities of talks characterized by the economics terms employed, hence on the bases of their economic content similarity. From such an investigation three main results are derived:
\begin{itemize}
\item The US Presidents speeches share the use of a core dictionary referred to economics and finance, see Figure \ref{fig:wordclouds} to see some examples.
\item The speeches clusterization based on the cosine similarity network lead to a core-periphery structure. Namely, the 948 analysed speeches are divided in two sets, one made by stronger connected speeches and another with lighter edges.
\item The words that lead the division in a core-periphery structure are those peculiar of certain events, namely the terms used to explain some local phenomena, see the differences between the locutions present in the bottom clouds of Figures \ref{fig:wordclouds}.
\end{itemize}

Concluding, the analysis of Figure \ref{crisisdummy} and the comments to the network clustering indicators provide clues about the potential explanations for such a core-periphery structure. Indeed, we consider plausible to hypothesize that the core-periphery setup is linked to the presence of two regimes: one for the speeches stated in the older critical periods and another made by speeches more recently stated during non-recession periods.

These results throw the basis for further researches. For example, the causes for the utilization of certain terms might be further investigated and they might confirm the connections with events like crises. The timing for the presence of some terms can be taken into consideration as well. Indeed, the combination of times and terminology resulted to be the main elements related to a core-periphery structure of the network. 
Finally, the sentiments associated with the terms of the economics glossary can be object of study. We believe that in this way it would be possible to determine the impact of the Presidents' party affiliation on the network.

\clearpage


\bibliographystyle{model5-names}
\biboptions{authoryear}

\end{document}